\documentclass[aps,prb,twocolumn,superscriptaddress,floatfix,longbibliography]{revtex4-2}

\usepackage{graphicx}
\usepackage{amsmath,amssymb,bm}
\usepackage{siunitx}
\usepackage[colorlinks=true,citecolor=blue,linkcolor=blue,urlcolor=blue]{hyperref}

\newcommand{\nv}{\bm{n}}
\newcommand{\Dc}{\bm{D}_{\mathrm{c}}}
\newcommand{\ec}{\varepsilon_{\mathrm{c}}}
\newcommand{\kB}{k_{\mathrm{B}}}
\newcommand{\Ms}{M_{\mathrm{s}}}
\newcommand{\zh}{\hat{\bm{z}}}
\newcommand{\Heff}{\bm{H}_{\mathrm{eff}}}
\newcommand{\Dfree}{D_{\mathrm{free}}}
\newcommand{\Dav}{D_{\mathrm{av}}}

\newsavebox{\figonebox}
\newcommand{\figonelabel}[1]{\makebox[1.35em][l]{\raisebox{\dimexpr\ht\figonebox-6.5pt\relax}{%
  \fontfamily{ptm}\fontsize{8}{9}\selectfont(#1)}}}

\begin{document}

\title{Steering Skyrmion Diffusion with Chiral Molecules}

\author{Palina Shaban}
\email{palina.shaban@ista.ac.at}
\affiliation{Institute of Science and Technology Austria (ISTA),
             Am Campus 1, 3400 Klosterneuburg, Austria}

\author{Ulrich Pototschnig}
\affiliation{Department of Chemistry, Universität Hamburg, Luruper Chaussee 149, 22791 Hamburg, Germany}
\author{Gulnaz Rakhmanova}
\affiliation{Institute of Science and Technology Austria (ISTA),
             Am Campus 1, 3400 Klosterneuburg, Austria}
\author{Mikhail Lemeshko}
\affiliation{Institute of Science and Technology Austria (ISTA),
             Am Campus 1, 3400 Klosterneuburg, Austria}

\author{Ragheed Alhyder}
\email{ragheed.alhyder@ista.ac.at}
\affiliation{Institute of Science and Technology Austria (ISTA),
             Am Campus 1, 3400 Klosterneuburg, Austria}

\begin{abstract}
Chiral molecules adsorbed on a magnetic film interact with its magnetization and can thereby alter the thermal motion of magnetic skyrmions. Here we develop a theoretical model in which an aligned layer of tilted molecules modifies the interfacial magnetic interactions according to its handedness. Within this model a skyrmion responds only to spatial variations of the coupling, so a uniform layer leaves its position unaffected, whereas boundaries and coverage gradients exert forces on it. Near a chiral region, these forces deflect the skyrmion transversely in a direction that reverses between enantiomers, which provides a direct signature of the chiral interaction. In extended periodic patterns, the same forces reshape the long-time diffusion, suppressing it in the direction in which the potential varies and enhancing it in the perpendicular direction. The net effect on the mean diffusion is set by the balance between the gyrotropic and dissipative responses of the skyrmion, as we show exactly for one-dimensional landscapes. Stochastic micromagnetic simulations and collective-coordinate calculations confirm these predictions and show how molecular orientation, spatial patterning, and skyrmion properties together determine the motion.
\end{abstract}

\maketitle

\section{Introduction}\label{sec:intro}

Magnetic skyrmions are localized spin textures whose nonzero topological charge counts
the winding of the magnetization across the film, and they behave as stable, particle-like quasiparticles in magnetic materials
\cite{skyrme1961,bogdanov1989,muhlbauer2009,yu2010,heinze2011}. In thin layers,
exchange, anisotropy and Dzyaloshinskii--Moriya interactions stabilize them at sizes
from a few to hundreds of nanometers \cite{sampaio2013,woo2016,moreauluchaire2016,boulle2016},
which makes them candidates for low-power storage, logic and spintronic devices
\cite{Mishra2025,Finocchio2016,Dohi2022}. Because the texture stays well defined while
its center moves, the dynamics reduce to those of an effective particle
\cite{nagaosa2013,fert2017,back2020,everschor2018} that responds to currents, fields,
interfaces and material inhomogeneities \cite{fert2013,iwasaki2013}. At finite
temperatures, the motion is Brownian and can be observed at room temperature
in weakly pinned films \cite{zazvorka2019,zhao2020,nozaki2019,dohi2023}, leading to proposals for using thermal
skyrmion motion as a resource for
stochastic and probabilistic computing \cite{prychynenko2018,pinna2018,raab2022}.

The response of a skyrmion to a force differs qualitatively from that of an overdamped
particle. Its topological charge adds a gyrotropic term to the collective-coordinate
dynamics, which rotates the velocity away from the force
\cite{thiele1973,everschor2012,makhfudz2012}. Under driving, this rotation causes the
skyrmion Hall effect \cite{jiang2017,litzius2017}, and in thermal motion it gives the
long-time diffusion constant a characteristic dependence on the Gilbert damping
\cite{schutte2014,miltat2018,weissenhofer2020}. Confinement can partially lift the
resulting gyrotropic suppression, so defects and interactions can enhance diffusion
\cite{rieger2023,schick2024}. Controlling skyrmion diffusion therefore requires control
over both the energy landscape and how its gradients are converted into motion.

Chiral molecular layers offer a way to engineer such structured
magnetic environments. Chiral molecules exhibit spin-selective
transport and polarization effects, collectively known as
chiral-induced spin selectivity
\cite{gohler2011,naaman2012,naaman2019,bloom2024}, and
adsorbed layers can modify the magnetization of a ferromagnet
without an applied current \cite{bendor2017}.
Theoretically, molecular dipoles can modify magnetic anisotropy
through interfacial spin-orbit coupling \cite{alhyder2023},
while tilted chiral molecular potentials can produce
handedness-dependent Hall responses \cite{alhyder2025}
and spin-polarized Josephson currents \cite{kuliashov2026}.
More recently, experiments on CoFeB have shown that chiral
polypeptides affect skyrmion stability and thermal dynamics
\cite{kapon2025}, while nitrogen-vacancy magnetometry has
revealed changes in skyrmion size, spacing, and shape with
molecular handedness and field polarity \cite{hondamuni2026}.
These observations motivate identifying the interfacial
interactions that distinguish molecular enantiomers and
determining how they influence skyrmion motion.

Here we use the symmetry of interfacial DM interactions \cite{rohart2013,nembach2015,srivastava2018} to construct an enantiomer-odd response permitted by an aligned and tilted molecular layer. An axisymmetric skyrmion couples to the spatial curl of this response, so molecular boundaries and coverage variations generate a handed energy landscape, as illustrated in Fig.~\ref{fig:schematic}(a).
Gyrotropic motion converts the resulting force into a predominantly transverse deflection that reverses with molecular handedness. Extended patterns instead produce anisotropic diffusion, suppressing diffusion in the direction of the potential variation and enhancing it in the perpendicular direction. The balance between gyrotropic and dissipative motion determines how the mean diffusion changes. Handedness therefore controls the transient direction, while the long-time diffusion tensor is the same for the reflection-related patterns considered here. These observables test the mechanism without requiring a quantitative calibration of a particular molecular interface.

\section{Model}\label{sec:model}
\subsection{Magnetic energy}\label{sec:energy}
We describe the magnetic film by a unit vector field $\nv(\bm r)$ specifying the local
magnetization direction and assume that it is uniform across the film thickness $d$.
In the absence of adsorbates, the energy density is
\begin{equation}
  w_0 = A(\nabla\nv)^2 - K n_z^2 - \Ms B\,n_z + w_{\mathrm N},
  \label{eq:w0}
\end{equation}
where $A$ is the exchange stiffness, $K$ is the effective perpendicular anisotropy,
$\Ms$ is the saturation magnetization and $B$ is a perpendicular magnetic field that
we set to zero below.

The interfacial DM interaction $w_{\mathrm N}$ is linear in magnetization gradients and
selects a definite rotational sense. To describe the native interaction and the
molecular contribution in a common form, we start from the energy
$\bm d^{(\delta)}\cdot(\bm S_{\bm r}\times\bm S_{\bm r+a\hat{\bm\delta}})$ of a
nearest-neighbor bond along $\hat{\bm\delta}$ with DM vector $\bm d^{(\delta)}$, where
$\bm S_{\bm r}$ is the spin at lattice site $\bm r$ and $a$ is the lattice constant. For
a texture that varies slowly on the lattice scale, expanding
$\bm S_{\bm r+a\hat{\bm\delta}}$ to first order in $a$ and summing the bonds gives the
continuum density
\begin{equation}
  w_{\mathrm{DM}} =
  \sum_{\delta=x,y}
  \bm D^{(\delta)}\cdot
  \left(\nv\times\partial_\delta\nv\right),
  \label{eq:wdm}
\end{equation}
\begin{figure}[ht]
  \sbox{\figonebox}{\includegraphics[width=\dimexpr\columnwidth-1.35em\relax]{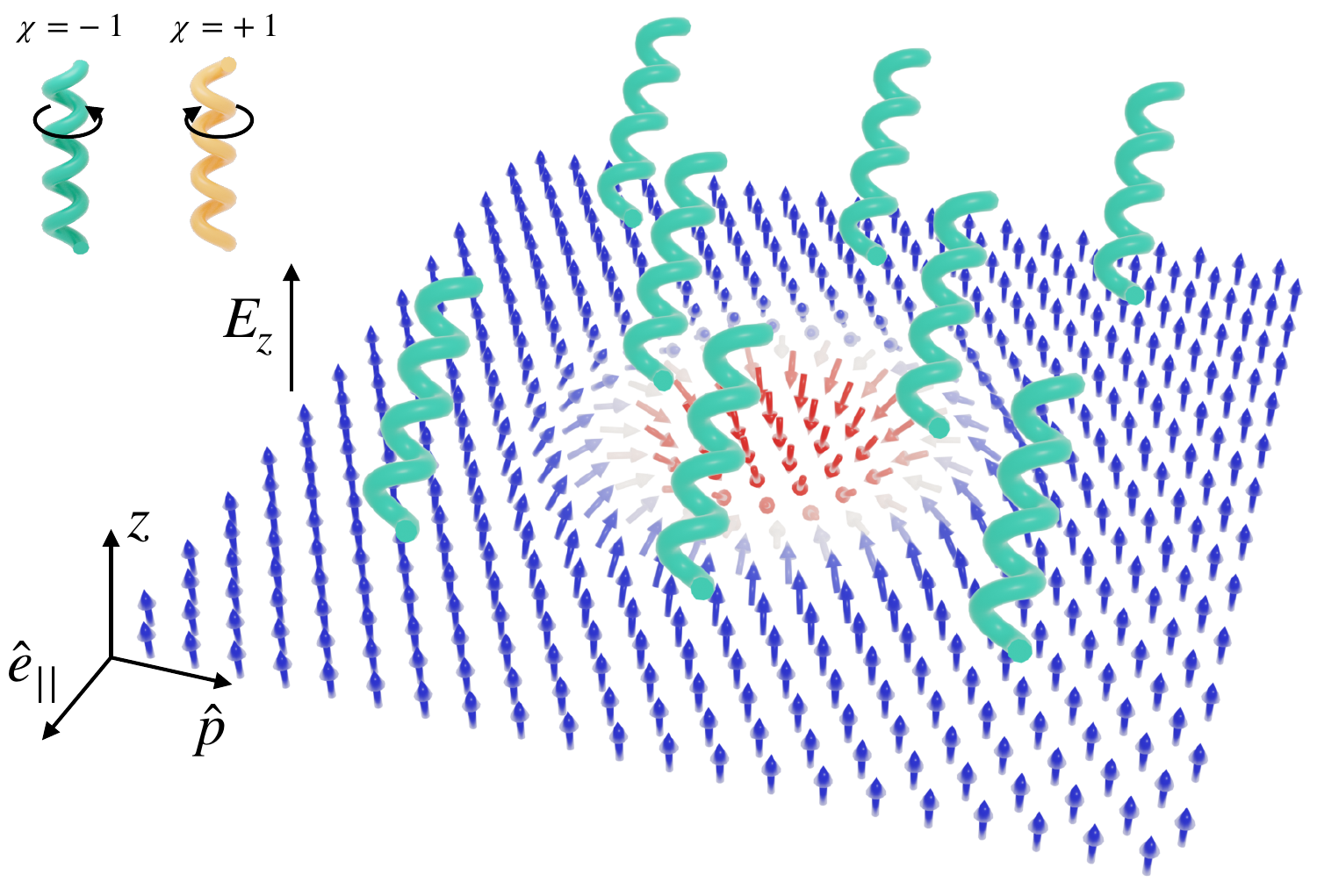}}%
  \figonelabel{a}\usebox{\figonebox}\\[3pt]
  \includegraphics[width=\columnwidth]{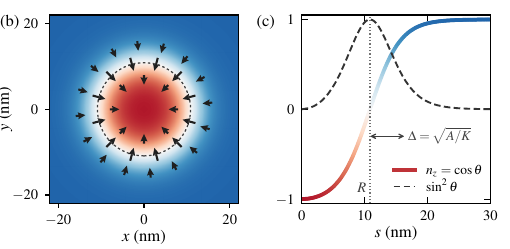}
  \caption{\label{fig:schematic}%
  Chiral molecular interface and N\'eel skyrmion.
  (a) Skyrmion beneath an aligned layer of tilted chiral molecules of handedness
  $\chi=\pm1$, with in-plane tilt direction $\hat{\bm p}$,
  $\hat{\bm e}_\parallel=\hat{\bm p}\times\zh$ and interfacial field $E_z$.
  (b) Relaxed reference skyrmion in the film plane $(x,y)$, with $n_z$ in color on the
  scale of (c), the in-plane magnetization as arrows and the radius $R$ dashed.
  (c) Azimuthal averages of $n_z=\cos\theta$ and $\sin^2\theta$ against the distance $s$
  from the skyrmion center, with $R$ at $n_z=0$ and $\Delta=\sqrt{A/K}$ for scale.}
\end{figure}%
where each vector $\bm D^{(\delta)}$ has units of energy per area and absorbs the spin
length and the lattice-to-continuum conversion. For a Rashba-type interfacial mechanism,
an effective polar field $\bm E$ gives
$\bm D^{(\delta)}=\lambda_{\rm so}\bm E\times\hat{\bm\delta}$
\cite{crepieux1998,kim2013,kundu2015}, where $\lambda_{\rm so}$ is an effective
spin-orbit coefficient. This form excludes a DM interaction of bulk type, which would
require $\bm D^{(\delta)}\parallel\hat{\bm\delta}$, because $\bm E\times\hat{\bm\delta}$
is always perpendicular to $\hat{\bm\delta}$. We use the relation only to fix the
geometry of the two channels below.

The original interface is represented by a field normal to the film,
$\bm E=E_z\zh$. The Rashba relation then gives
$\bm D^{(x)}=-D_0\hat{\bm y}$ and $\bm D^{(y)}=D_0\hat{\bm x}$, where
$D_0=-\lambda_{\rm so}E_z$ is the native DM constant. Substituting these vectors into Eq.~\eqref{eq:wdm}
gives the N\'eel DM energy \cite{rohart2013}
\begin{equation}
  w_{\mathrm{N}} = D_0\left[\, n_x\partial_x n_z + n_y\partial_y n_z
                              - n_z\nabla\!\cdot\!\nv \,\right].
  \label{eq:wN}
\end{equation}
Its competition with the anisotropy is measured by $\kappa=\pi D_0/(4\sqrt{AK})$. For
$\kappa<1$, the regime studied here, the uniform state is the ground state and an
isolated skyrmion is metastable. The resulting N\'eel skyrmion consists of a reversed
core surrounded by a domain wall in which the magnetization rotates through the film
plane, as the relaxed texture in Fig.~\ref{fig:schematic}(b) shows. We characterize its
size by the radius $R$, at which $n_z$ changes sign, and by the exchange-anisotropy
length $\Delta=\sqrt{A/K}$ that sets the scale of the wall. The anisotropy $K$ includes
the local part of the demagnetizing field, and we omit nonlocal magnetostatic
interactions.

An aligned layer of tilted chiral molecules defines a preferential
in-plane direction $\hat{\bm p}$, corresponding to the projection
of their molecular axes onto the film plane, as illustrated in
Fig.~\ref{fig:schematic}(a). For a molecular axis
$\hat{\bm m}$ tilted by an angle $\vartheta$ from the surface normal,
we write
$\hat{\bm m}=\cos\vartheta\,\zh+\sin\vartheta\,\hat{\bm p}$.
Together with the surface normal, $\hat{\bm p}$ defines a second
in-plane vector $\hat{\bm e}_\parallel=\hat{\bm p}\times\zh$,
perpendicular to the tilt. Since this cross product is an axial
vector and the molecular handedness $\chi=\pm1$ is a pseudoscalar,
their product transforms as a polar vector. An enantiomer-odd
contribution to the effective Rashba field can therefore
be written as
$\bm E_{\rm odd}=\chi E_\chi\hat{\bm e}_\parallel$.
At fixed molecular orientation, this contribution reverses
with handedness and vanishes upon averaging over uniformly
distributed in-plane tilt directions. The coefficient $E_\chi$
characterizes the interfacial spin-orbit response and depends
on microscopic bonding, molecular tilt, and orientational order.
We treat the resulting magnetic coupling phenomenologically,
without assuming a particular microscopic mechanism.
A tilted layer of either handedness can also produce an in-plane field
$E_p\hat{\bm p}$ along the tilt, which is the same for both enantiomers.
Through the same Rashba relation it adds a coupling vector along
$\hat{\bm e}_\parallel$. Section~\ref{sec:symmetry} shows that the skyrmion
responds to the curl of the coupling, which involves only coverage gradients
perpendicular to the coupling vector. This even contribution therefore acts on
gradients along $\hat{\bm p}$, whereas the odd one acts on gradients along
$\hat{\bm e}_\parallel$, and Sec.~\ref{sec:regime} uses this difference to tell
the two apart.

We retain this enantiomer-odd contribution and treat
enantiomer-even modifications of the magnetic interface
separately. Applying the Rashba relation to the in-plane field
$\bm E_{\rm odd}$ produces DM vectors normal to the film. Their coefficients
along the two in-plane bond directions define the vector
\begin{equation}
  \Dc(\chi)
  = \chi\lambda_{\rm so}E_\chi
    \big(\zh\times\hat{\bm e}_\parallel\big)
  = \chi\lambda_{\rm so}E_\chi\hat{\bm p}.
  \label{eq:Dchi}
\end{equation}
We choose the sign assigned to $\chi$ so that
$\lambda_{\rm so}E_\chi/D_0>0$, and write the relative coupling as
$\ec=\chi\eta$, where $\eta=|\Dc|/|D_0|$ is nonnegative.
When the two channels share the same spin-orbit coefficient, this convention gives
$E_\chi/E_z=-\eta$.

Substituting the corresponding bond vectors
$\bm D_{\rm c}^{(x)}=D_{{\rm c}x}\zh$ and
$\bm D_{\rm c}^{(y)}=D_{{\rm c}y}\zh$
into Eq.~\eqref{eq:wdm} gives
\begin{equation}
  w_{\mathrm c}
  = \Dc\cdot
    \left(n_x\nabla n_y-n_y\nabla n_x\right).
  \label{eq:wc}
\end{equation}
Unlike the original N\'eel interaction, this term couples
only to spatial rotations of the in-plane magnetization.
Adsorption may also modify the perpendicular magnetic anisotropy.
We account for this through an enantiomer-even contribution
$
    w_{\rm a} = A_{\rm a} n_z^2,\,
    A_{\rm a} = \beta |\Dc|^2/(4A),
$
where $\beta$ is a phenomenological parameter that can have either
sign. This correction is quadratic in the molecular coupling and
changes the effective anisotropy from $K$ to $K-A_{\rm a}$.
We use $\beta=1$ as a reference, corresponding to
$A_{\rm a}=0.3\,\ec^2 K$ at full coverage, and examine the
dependence on $\beta$ below. A smaller, higher-gradient correction
is included in the numerical implementation and discussed in
Appendix~\ref{app:numerics}.

In the calculation, the molecules therefore enter through the single vector field
\begin{equation}
  \Dc(\bm r)=\ec D_0\,f(\bm r)\,\hat{\bm p},
  \label{eq:Dr}
\end{equation}
where $f$ is the local coverage normalized to unit peak. The handedness is carried by
the sign of $\ec=\chi\eta$, so exchanging enantiomers at fixed tilt and coverage amounts
to $\ec\to-\ec$, while $A$, $K$, $D_0$ and $\Ms$ are the same for both. We take $\hat{\bm p}=\hat{\bm y}$, and hence
$\hat{\bm e}_\parallel=\hat{\bm x}$, throughout. We consider three coverage profiles, a
uniform layer, an isolated Gaussian patch representing a small group of molecules, and a
periodic array of such patches representing an extended pattern. Comparing them
separates the effect of the amount of coverage from that of its spatial structure. We
use $\eta\leq0.1$,
where the anisotropy correction $w_{\rm a}$ remains small compared with the odd channel.

The total energy density is $w=w_0+w_{\mathrm{c}}+w_{\mathrm{a}}$. Under $\ec\to-\ec$
only $w_{\rm c}$ changes, reversing its sign, because $w_0$ does not involve the
molecules and $w_{\rm a}$ is quadratic in $\ec$. The native coupling $D_0$ also keeps its
sign in this exchange, so a local change of $D_0$ by adsorption could alter the size and
energy of the skyrmion but could not distinguish the enantiomers. The translational
potential $U$, the energy of the rigid texture as a function of the position of its
center, therefore has the form $U(\ec)=\ec U_1+\ec^2U_2$ up to a constant, where $U_1$
and $U_2$ do not depend on the handedness. Its odd part
$U_{\mathrm{odd}}=[U(\ec)-U(-\ec)]/2=\ec U_1$ distinguishes the enantiomers, whereas the
even part $U_{\mathrm{even}}=[U(\ec)+U(-\ec)]/2$ is common to both. We report the two
parts separately.

\subsection{Stochastic dynamics}\label{sec:sllg}

The magnetization obeys the stochastic Landau--Lifshitz--Gilbert equation
\begin{equation}
  \partial_t\nv = -\frac{\gamma_0}{1+\alpha^2}\,\nv\times\bm{H}
                  -\frac{\gamma_0\alpha}{1+\alpha^2}\,\nv\times(\nv\times\bm{H}),
  \label{eq:llg}
\end{equation}
where $\gamma$ is the gyromagnetic ratio, $\mu_0$ is the vacuum permeability,
$\gamma_0=\gamma\mu_0$ and $\alpha$ is the Gilbert damping. The field
$\bm{H}=\Heff+\bm{H}_{\mathrm{th}}$ is the sum of the effective and thermal fields.
The effective field is $\Heff=-(\mu_0\Ms d)^{-1}\delta E/\delta\nv$, where $E=d\int w\,d^2r$ is the energy of
the film. We compute it as the exact derivative of the discretized energy rather than by
discretizing the continuum expression. This matters when $\Dc$ and $A_{\rm a}$ vary in
space, because the derivative then contains gradients of the couplings themselves (see Appendix~\ref{app:numerics}). The thermal
field is Gaussian and white in space and time, and the fluctuation-dissipation theorem
fixes its amplitude \cite{brown1963,kubo1966},
\begin{equation}
  \left\langle H^a_{i}(t)\,H^b_{j}(t')\right\rangle
  = \frac{2\alpha\kB T}{\gamma\mu_0^2\Ms V}\,
    \delta_{ab}\,\delta_{ij}\,\delta(t-t'),
  \label{eq:fdt}
\end{equation}
where $a$ and $b$ label Cartesian components of the thermal field, $i$ and $j$ label
mesh cells of volume $V$, $\kB$ is the Boltzmann constant and $T$ is the temperature.

We integrate Eq.~\eqref{eq:llg} at $T=\SI{300}{K}$ unless stated otherwise, using the
Stratonovich form of the Heun scheme with the same noise sample in the predictor and the
corrector \cite{garciapalacios1998,mentink2010}. The time step resolves the stiff
exchange modes and is reduced further at low damping to control the stochastic
integration error. We test thermal equilibrium against the exact Boltzmann distribution
of a macrospin and against the magnetization of the extended film, with values and
tolerances given in Appendix~\ref{app:numerics}.

The topological charge and the skyrmion position are both obtained from the density
$q=\nv\cdot(\partial_x\nv\times\partial_y\nv)$, evaluated with central differences. The
charge $Q=(4\pi)^{-1}\int q\,d^2r$ is $-0.99$ for the relaxed texture and stays near that value during the simulation. The
position is a circular mean of the coordinate weighted by the signed density $q$.

\subsection{Reduction to the motion of a point}\label{sec:thiele}

For transport we only need the position of the skyrmion, and the full magnetization
dynamics reduces to it when the texture moves without changing shape. We set
$\nv(\bm{r},t)=\nv_0(\bm{r}-\bm{R}(t))$, where $\nv_0$ is the relaxed texture and
$\bm R(t)$ is the position of its center. Projecting Eq.~\eqref{eq:llg} onto the
translational mode then gives an equation for $\bm{R}$ alone \cite{thiele1973},
\begin{equation}
  \frac{\Ms d}{\gamma}\left[\,G\,\zh\times\dot{\bm{R}}
        + \alpha\mathcal{D}\,\dot{\bm{R}}\,\right] = \bm{F}+\bm\zeta ,
  \label{eq:thiele}
\end{equation}
where $G$ and $\mathcal D$ are the gyrotropic and dissipative coefficients discussed
below, $\bm F=-\nabla_{\bm R}U$ is the force from the energy $U(\bm R)$ of the texture
centered at $\bm R$, and the thermal force $\bm\zeta$ has covariance
$\langle\zeta_i(t)\zeta_j(t')\rangle=2\alpha\mathcal D(\Ms d/\gamma)\kB T
\delta_{ij}\delta(t-t')$. In a clean film $U$ is flat and only the thermal force acts,
whereas spatial variations of $\Dc$ make $U$ position dependent and add the force
$\bm F$. We use this rigid-texture approximation for long-time transport, knowing that it neglects internal modes, thermally
induced deformations and interactions between skyrmions.

The second term on the left of Eq.~\eqref{eq:thiele} is ordinary friction. Its
coefficient $\mathcal{D}=\int(\partial_x\nv)\cdot(\partial_x\nv)\,d^2r$ is a
dimensionless measure of the magnetization gradients and generally grows with the
circumference of the skyrmion wall. The first term acts perpendicular to the velocity,
survives as $\alpha\to0$ and does no work, like the magnetic force on a moving charge.
Its dimensionless coefficient $G=4\pi Q$ is fixed by the topological charge, and the
material enters both terms only through the common prefactor $\Ms d/\gamma$. For the
reference skyrmion of Sec.~\ref{sec:baseline}, $\mathcal{D}=17.6$ and $Q=-0.99$, so at
$\alpha=0.1$ the ratio $|G|/(\alpha\mathcal{D})\simeq7.1$ and the gyrotropic term
dominates friction by nearly an order of magnitude.

Inverting Eq.~\eqref{eq:thiele} shows that a force along one direction produces a
velocity rotated from it by nearly a right angle. Its components $v_\parallel$ along
the force and $v_\perp$ perpendicular to it are in the ratio
\begin{equation}
  \left|\frac{v_\perp}{v_\parallel}\right| = \frac{|G|}{\alpha\mathcal{D}} .
  \label{eq:hall}
\end{equation}
A skyrmion pushed by a chiral patch therefore moves almost perpendicular to the push.
The adsorption geometry sets the axis of the force and the handedness sets its sign,
whereas the angle of the deflection is set by the magnetic texture rather than by the
molecular chemistry. The same inversion, together with the fluctuation-dissipation
theorem, gives the free diffusion constant
\begin{equation}
  \Dfree = \frac{\kB T\gamma}{\Ms d}\,
      \frac{\alpha\mathcal{D}}{G^2+(\alpha\mathcal{D})^2},
  \label{eq:Ddiff}
\end{equation}
which serves as the baseline for all simulations below
\cite{schutte2014,miltat2018,weissenhofer2020}. For $|G|\gg\alpha\mathcal D$ it reduces
to $\Dfree\simeq\kB T\gamma\alpha\mathcal D/(\Ms dG^2)$, which grows with the damping.
The thermal force vanishes as $\alpha\to0$, while the gyrotropic response that converts
it into motion stays finite, and the massless description has no cyclotron orbit. For a
patterned film we report the long-time diffusion tensor $D_{ij}$, defined by
$\langle\Delta R_i\Delta R_j\rangle\simeq2D_{ij}t$ for the displacement $\Delta\bm R$
during a time $t$, with all coefficients normalized by $\Dfree$. Its mean
$\Dav=(D_{xx}+D_{yy})/2$ fixes the displacement variance $C(t)\simeq4\Dav t$.

\subsection{Spatial coupling and symmetry}
\label{sec:symmetry}

We now determine the potential generated by the molecular
coupling of Eq.~\eqref{eq:Dr}. For a rigid skyrmion centered
at $\bm R$, its enantiomer-odd part is
$U_{\rm odd}(\bm R)=d\int w_{\rm c}\,d^2r$.
An axisymmetric texture can be written as
$n_z=\cos\theta(s)$ and
$\nv_\parallel=\sin\theta(s)(\cos\phi,\sin\phi)$,
where $\phi=\varphi+\psi$, with $s$ and $\varphi$ the polar
coordinates relative to $\bm R$ and $\psi$ the helicity.
The interaction of Eq.~\eqref{eq:wc} then becomes
$
    w_{\rm c}
    = \sin^2\theta(s)
      \Dc\cdot\hat{\bm\varphi}/s,
$
where $\hat{\bm\varphi}=(-\sin\varphi,\cos\varphi)$.
The coupling is independent of helicity and measures
the circulation of $\Dc$ around the skyrmion.
For a uniform field this circulation vanishes, so a
uniform molecular layer exerts no force on a rigid
axisymmetric skyrmion.

The weight $\sin^2\theta(s)$ is concentrated in the
skyrmion wall, as shown in Fig.~\ref{fig:schematic}(c).
For a coupling that varies slowly across the texture,
expanding $\Dc$ about $\bm R$ and integrating over
the azimuth gives
\begin{equation}
    U_{\rm odd}(\bm R)\simeq
    \pi d\int_0^\infty ds\,s\sin^2\theta(s)\,
    [\partial_xD_{{\rm c}y}
    -\partial_yD_{{\rm c}x}]_{\bm R}.
    \label{eq:curl}
\end{equation}
Thus, the odd potential is governed by the spatial curl
of the molecular DM coupling, rather than its local
magnitude. This holds exactly for a rigid axisymmetric
texture, because by Stokes' theorem the line integral of
$\Dc$ around a circle of radius $s$ centered at $\bm R$
equals the flux of $\nabla\times\Dc$ through the enclosed
disk. The local expression above further assumes that
the curl varies slowly across the skyrmion, and at
shorter length scales the full convolution with the
skyrmion texture must be retained.

For $\Dc=\ec D_0 f(\bm r)\hat{\bm p}$, with
$\hat{\bm p}=\hat{\bm y}$, the curl reduces to
$\ec D_0\partial_x f$, where $\hat{\bm x}
=\hat{\bm e}_\parallel$. A Gaussian molecular patch
therefore creates an attractive and a repulsive lobe
on opposite sides of the patch, perpendicular to the
molecular tilt. Exchanging enantiomers reverses these
lobes, whereas the even anisotropy contribution is
unchanged. The result is consistent with reflection
in the plane spanned by $\zh$ and $\hat{\bm p}$,
which reverses both the handedness $\chi$ and the
skyrmion charge $Q$ while preserving the energy
(Appendix~\ref{app:numerics}). Reversing $Q$ alone
at zero field leaves the potential unchanged but
reverses the gyrotropic response and hence the
transverse deflection.

These forces do not produce sustained directed motion
in equilibrium. The stationary distribution is
proportional to $\exp[-U(\bm R)/(\kB T)]$, and
gyrotropy generates only circulating probability
currents, with no net flux through a periodic cell.
Molecular handedness therefore controls the transient
deflection following a change in the coupling, rather
than a steady drift. The symmetric long-time diffusion
tensor is even under $G\to-G$ at fixed potential.
For the reflection-related patterns considered below,
it is also identical for the two enantiomers.

\section{Results}\label{sec:results}

\subsection{Free diffusion and the chiral selection rule}\label{sec:baseline}

Unless stated otherwise, we use $A=\SI{1.5e-11}{J/m}$, $K=\SI{5e5}{J/m^3}$,
$D_0=\SI{3e-3}{J/m^2}$, $\Ms=\SI{1.1e6}{A/m}$ and $d=\SI{5}{nm}$. These values give
$\kappa=0.86$ and are representative of CoFeB, but they define an idealized thin-film
model rather than a calibration of the experimental stack. On the \SI{1}{nm} mesh the
relaxed isolated skyrmion of Fig.~\ref{fig:schematic}(b) has radius \SI{10.9}{nm},
charge $-0.99$ and $\mathcal D=17.6$. Section~\ref{sec:regime} lowers $K$ to obtain a
larger texture while keeping all other parameters fixed.

\begin{figure}[t]
  \includegraphics[width=\columnwidth]{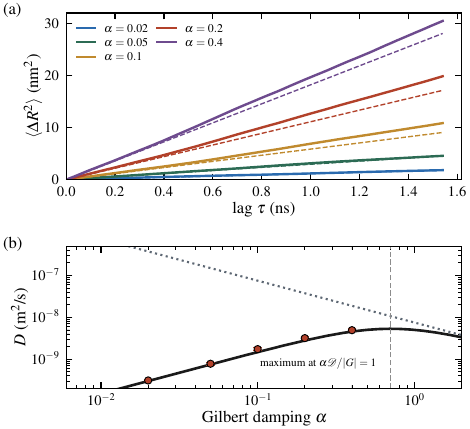}
  \caption{\label{fig:baseline}%
  Free skyrmion diffusion in the reference film.
  (a) Connected mean squared displacement $\langle\Delta R^2\rangle$ of the skyrmion
  center against lag $\tau$ for several Gilbert damping values $\alpha$, from stochastic
  LLG simulations (solid) and the Thiele prediction $4\Dfree\tau$ (dashed).
  (b) Diffusion constant $D$ against $\alpha$ from LLG (points), Eq.~\eqref{eq:Ddiff}
  (solid) and the dissipative limit $G=0$ (dotted). The dashed line marks the maximum of
  Eq.~\eqref{eq:Ddiff} at $\alpha\mathcal D/|G|=1$.}
\end{figure}

We first follow the reference skyrmion in a clean film, without molecules, using
stochastic LLG simulations at several values of the damping, and compare the result
with the Thiele description of Sec.~\ref{sec:thiele}. The mean
squared displacement in Fig.~\ref{fig:baseline}(a) is linear in lag, with a slope that
grows with $\alpha$~\cite{schutte2014}. The massless collective-coordinate model misses only the short-time
bend of the data, which comes from cyclotron motion, frequency-dependent inertia and
gyrodamping and does not affect the asymptotic diffusion constant.
Figure~\ref{fig:baseline}(b) collects the diffusion constants fitted to the long-time
slopes and compares them, over a wider range of damping, with Eq.~\eqref{eq:Ddiff} and
with the purely dissipative value obtained by setting $G=0$. The fitted constants agree
with the Thiele values within three to eighteen percent for
$0.02\leq\alpha\leq0.4$, and each deviation is consistent with its statistical
uncertainty. At $\alpha=0.1$ the gyrocoupling reduces the Thiele value by the factor
$1+G^2/(\alpha\mathcal D)^2\simeq51$ relative to the dissipative value.
Equation~\eqref{eq:Ddiff} has its maximum at $\alpha\mathcal D/|G|=1$, and all our
calculations lie below this turnover, where increasing the damping increases the
diffusion (see Appendix~\ref{app:numerics} for the numerical checks).

\label{sec:res-curl}

We test Eq.~\eqref{eq:curl} by translating the relaxed texture across several coupling
profiles and evaluating its energy. A uniform chiral field produces no position-dependent
energy to numerical precision. A localized patch produces an attractive and a repulsive
side, which exchange when $\ec$ is reversed, with a peak-to-peak energy that grows
linearly in $|\ec|$ and reaches about $8\,\kB T$ at $|\ec|=0.1$. Periodic and ramped
profiles, which also have nonzero curl, act on the skyrmion as well. Spatial variation is
therefore the source of the force, and uniform coverage is inert at linear order.

\subsection{Motion in structured chiral layers}\label{sec:drift}

\begin{figure*}[t]
  \includegraphics[width=\textwidth]{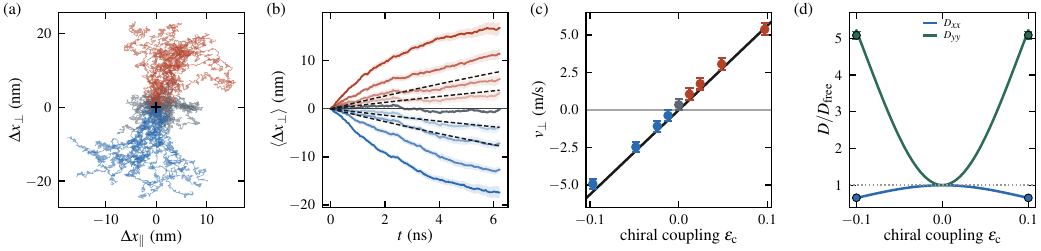}
  \caption{\label{fig:drift}\label{fig:tensor}%
  Skyrmion motion in chiral adsorbate potentials.
  (a)--(c) Transient response to a Gaussian patch of width $\sigma=\SI{10}{nm}$ switched
  on at $t=0$, at $\alpha=0.1$, for $\ec>0$ (red), $\ec<0$ (blue) and the reference film
  (gray).
  (a) Trajectories of the skyrmion center, 16 per condition at $|\ec|=0.0969$, with
  displacements $\Delta x_\parallel$ along $\hat{\bm e}_\parallel$ and $\Delta x_\perp$
  along $\zh\times\hat{\bm e}_\parallel$. The cross marks the patch center, where all
  trajectories start.
  (b) Mean transverse displacement $\langle\Delta x_\perp\rangle$ against time $t$ for
  $|\ec|=0.0121$, $0.0242$, $0.0484$ and $0.0969$, darker for larger $|\ec|$, with
  standard errors shaded. Dashed lines give the weak-coupling prediction of
  Eq.~\eqref{eq:green} for the two smallest couplings.
  (c) Mean transverse velocity $v_\perp$ over the first \SI{1.5}{ns} against $\ec$, with
  the initial-force Thiele prediction (line).
  (d) Long-time diffusion coefficients $D_{xx}$ and $D_{yy}$ of the periodic array at
  $\alpha=0.2$, from homogenization of the micromagnetic
  landscapes (lines) and from $2\times256$ collective-coordinate trajectories of
  \SI{6}{\micro s} (points). The dotted line marks the free film.}
\end{figure*}

A localized patch converts molecular handedness into a transient deflection. We place
the relaxed clean-film skyrmion at the center of a Gaussian patch of width \SI{10}{nm}
and switch on the coupling at $t=0$, with $\alpha=0.1$. Figure~\ref{fig:drift}(a)
shows the resulting trajectories for both enantiomers and for the reference film. Because the odd potential crosses zero at
the patch center with a finite slope, the initial force there is finite, and the two
enantiomers move to opposite sides while the reference skyrmion only wanders thermally.
Figure~\ref{fig:drift}(b) shows the mean transverse displacement and its growth with the
coupling. Figure~\ref{fig:drift}(c) shows the mean transverse velocity over the first
\SI{1.5}{ns}. Its antisymmetric part is linear in $\ec$ and agrees with the initial-force Thiele
prediction within one to six percent for all four coupling strengths. Repeating
these calculations at zero temperature separates the response to the force from the
thermal motion.

In the simulations the coupling is switched on at $t=0$, but the same displacement
appears without any switching. In the Thiele equation the future motion depends only on
the present position, so a skyrmion that starts at $\bm R_0$ in a film already covered
by molecules moves on average as in Fig.~\ref{fig:drift}(b). Tracking experiments can
therefore measure this displacement by averaging trajectory segments that start at the
same position. For weak coupling it follows from the free motion of the skyrmion,
\begin{equation}
 \langle\Delta\bm R(t)\rangle=\int_0^t ds\int d^2r\,G_0(\bm r-\bm R_0,s)\,
 \mathsf M\bm F(\bm r),
 \label{eq:green}
\end{equation}
where $G_0$ is the probability that a free skyrmion moves from $\bm R_0$ to $\bm r$ in
a time $s$, a Gaussian with diffusion coefficient $\Dfree$, and $\mathsf M$ is the
mobility obtained by inverting Eq.~\eqref{eq:thiele}. At short times this reduces to
the initial-force Thiele prediction. For $|\ec|\leq0.024$ it follows the micromagnetic
displacement over the full \SI{6.2}{ns} within two standard errors, as the dashed lines
in Fig.~\ref{fig:drift}(b) show, whereas for stronger coupling the skyrmion moves
beyond the region where the force is nearly constant. Since the force of a localized
patch integrates to zero, the displacement eventually saturates. For $|\ec|=0.0121$ it
reaches about \SI{35}{nm}, three skyrmion radii, half of it within about \SI{50}{ns}.
                                         
A check at zero temperature, with the skyrmion placed at six positions across a
periodic profile, confirms the mobility itself. The measured velocities follow the prediction with slopes of $1.006$ and $0.95$
for the transverse and longitudinal components, with relative residuals of $0.06$ and
$0.24$, the larger residual belonging to the weaker longitudinal component. In the
gyrotropic limit the signed transverse response is $v_\perp\simeq-\gamma F/(\Ms dG)$
along $\zh\times\hat{\bm F}$, where $F=|\bm F|$. The handedness thus sets the direction of the force, while
the topological charge sets the sense in which it is converted into transverse motion. At
$\eta\simeq0.024$ the two enantiomers separate by about \SI{13.3}{nm} after
\SI{6.2}{ns}, compared with \SI{5.9}{nm} of root-mean-square thermal wandering in the
reference film.

\label{sec:tensor}
An isolated patch changes only the transient motion, whereas an extended pattern can
modify the asymptotic diffusion. We therefore consider a periodic array of same-handed
Gaussian patches with a common tilt, a period of \SI{21.3}{nm} and a width of $0.235$
times the period. We obtain its translational landscape directly from the micromagnetic
energy of the rigid texture. At $\ec=0.1$ the landscape varies by $2.06\kB T$ from peak
to trough, compared with $0.18\kB T$ for the even part alone, and its one-dimensional
projection carries $99.4$ percent of the variance.
The resulting skyrmion potential varies predominantly along $x$ and is nearly uniform
along $y$. We therefore distinguish diffusion along the potential variation, $D_{xx}$,
from diffusion perpendicular to it, $D_{yy}$.

For a landscape that depends only on the component $X$ of $\bm R$ along $\hat{\bm x}$,
inverting Eq.~\eqref{eq:thiele} gives
$\dot X=-\mu U'(X)+\xi_X$ and $\dot Y=\nu U'(X)+\xi_Y$ with the dissipative and
gyrotropic mobilities
$\mu=\gamma\alpha\mathcal D/[\Ms d\,(G^2+(\alpha\mathcal D)^2)]$ and
$\nu=\gamma G/[\Ms d\,(G^2+(\alpha\mathcal D)^2)]$, so that $\nu/\mu=G/(\alpha\mathcal D)$,
and with thermal velocities obeying
$\langle\xi_i(t)\xi_j(t')\rangle=2\Dfree\delta_{ij}\delta(t-t')$. Motion along
$x$ is hindered by the potential barriers, while gyrotropy generates an additional
velocity along $y$ that follows the local potential gradient. This velocity vanishes on average in
equilibrium, but its temporal correlations enhance the spreading. Periodic
homogenization gives the exact result
\begin{equation}
 \frac{D_{xx}}{\Dfree}=L,\qquad
 \frac{D_{yy}}{\Dfree}=1+r^2(1-L),\qquad D_{xy}=0,
 \label{eq:tensor}
\end{equation}
where $r=|G|/(\alpha\mathcal D)$ and
$L=[\langle e^{U/\kB T}\rangle\langle e^{-U/\kB T}\rangle]^{-1}$ is the Lifson--Jackson
factor, with uniform averages over one period \cite{lifson1962}. The normalized
suppression along $x$ is independent of $r$, although $\Dfree$ itself depends
on $G$. The enhancement along $y$ is the finite-barrier counterpart of
confinement-enhanced gyrotropic diffusion. Exchanging $U$ and $-U$ leaves $L$ unchanged,
and the even adsorption term keeps the diagonal diffusivities equal for the
reflection-related profiles studied here, so the tensor is the same for both
enantiomers.

Landscapes are already known to enhance skyrmion diffusion, since defects and mutual
interactions raise the long-time diffusion at low damping, where the gyrotropic term
dominates \cite{rieger2023,schick2024}. This work identifies what an axisymmetric
skyrmion couples to, namely the curl of an enantiomer-odd interfacial channel rather than
the coverage itself, and gives the resulting transport as a closed-form tensor. The
response is therefore anisotropic rather than scalar. The mean diffusion rises or falls
according to whether $r$ exceeds one, independently of the landscape, and the magnitude
of the change is controlled by a pattern length scale set by the skyrmion form factor.

Figure~\ref{fig:tensor}(d) shows the two diffusion coefficients of the array against
$\ec$ at $\alpha=0.2$. They are computed by periodic homogenization of the full
two-dimensional landscape (Appendix~\ref{app:cell}) and checked against stochastic
collective-coordinate trajectories. At $\ec=0.1$ the homogenization gives
$D_{xx}/\Dfree=0.6534$ and $D_{yy}/\Dfree=5.2139$, in agreement with the stochastic
estimates $0.655\pm0.014$ and $5.09\pm0.10$. The one-dimensional result, $0.6516$ and $5.3690$,
differs by about $0.3$ and $3$ percent, a projection error that is separate from the
statistical uncertainty of the trajectories. The two-dimensional values agree between
the enantiomers and are stable under refinement of the cell solver. Diffusion
perpendicular to the potential variation is almost eight times faster than along it, and the mean diffusion is $2.934$
times the free value, a $193$ percent enhancement. Varying the independent anisotropy
coefficient over $-1\leq\beta\leq2$ changes the directional coefficients by less than
four percent and preserves this anisotropy. The effect is therefore governed by the odd
interfacial coupling rather than by the particular even correction.

\subsection{Enhanced and suppressed diffusion}\label{sec:regime}

From Eq.~\eqref{eq:tensor}, $\Dav/\Dfree-1=(r^2-1)(1-L)/2$. For any nonflat
one-dimensional potential, the mean diffusion therefore increases when gyrotropic motion
dominates and decreases when dissipation dominates, with the crossover exactly at $r=1$.
A two-dimensional landscape can shift this crossover, so its position must be calculated
rather than assumed.

\begin{figure}[t]
  \includegraphics[width=\columnwidth]{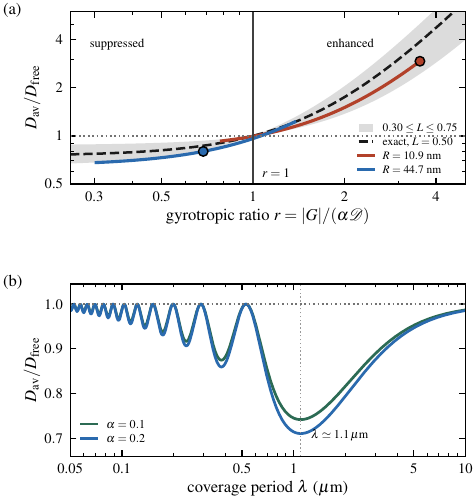}
  \caption{\label{fig:regime}%
  Control of the mean diffusion by damping and pattern period.
  (a) $\Dav/\Dfree$ against the gyrotropic ratio $r=|G|/(\alpha\mathcal D)$, from
  Eq.~\eqref{eq:tensor} for $L=0.50$ (dashed) and $0.30\leq L\leq0.75$ (band), and from
  two-dimensional homogenization (solid) for the skyrmions of radius \SI{10.9}{nm} and
  \SI{44.7}{nm} in coverage periods of \SI{21.3}{nm} and \SI{48}{nm}, with
  $0.2\leq\alpha\leq0.9$, $\eta=0.1$ and $\beta=1$. Circles mark $\alpha=0.2$ for the
  smaller and $\alpha=0.4$ for the larger skyrmion.
  (b) Thin-wall prediction of $\Dav/\Dfree$ against the period $\lambda$ of the coverage
  $f(X)=[1+\cos(2\pi X/\lambda)]/2$, for $R=\SI{322}{nm}$, $\Delta=\SI{6}{nm}$,
  $d=\SI{0.9}{nm}$, $D_0=\SI{3}{mJ/m^2}$, $\eta=0.1$ and $T=\SI{300}{K}$. The vertical
  line marks the strongest coupling, near $\lambda=3.41R$.
  Dotted horizontal lines mark the free film.}
\end{figure}

Figure~\ref{fig:regime}(a) plots $\Dav/\Dfree$ against $r$, from this result over a range
of $L$ and from the full micromagnetic landscapes of two textures, each followed as the
damping varies. The first is the reference skyrmion. The second is a larger texture,
obtained by lowering the anisotropy to $K=\SI{3.77625e5}{J/m^3}$, which has $R\simeq\SI{44.7}{nm}$,
$\mathcal D\simeq45.8$ and $Q\simeq-0.999$ on the \SI{0.5}{nm} mesh. Together they
establish both regimes. In a pattern of
period \SI{48}{nm}, the larger texture gives $\Dav/\Dfree=1.222$ at $\alpha=0.2$ and
$0.800$ at $\alpha=0.4$. Changing only the damping thus turns a $22$ percent enhancement
into a $20$ percent suppression, and the residual two-dimensional structure shifts the
crossover from the projected $\alpha=0.274$ to $0.258$. These results describe transport
within the rigid collective-coordinate model, not a finite-temperature micromagnetic
simulation of the larger texture, and Appendix~\ref{app:numerics} reports the mesh and
relaxation checks.

The size dependence comes from the wall. For an axisymmetric texture
$\mathcal D=\pi\int_0^\infty ds\,s[(\partial_s\theta)^2+\sin^2\theta/s^2]$, which for a
circular wall of width $\Delta=\sqrt{A/K}$ reduces to
$\mathcal D\simeq2\pi(R/\Delta+\Delta/R)$, and for $R\gg\Delta$ this gives
$r\simeq2|Q|\Delta/(\alpha R)$. The gyrotropic coefficient is fixed by the topological
charge, whereas the dissipation grows with the length of the wall, so larger skyrmions
approach the dissipative regime more easily. Figure~\ref{fig:scaling} in
Appendix~\ref{app:numerics} compares this law with the dissipation integral of textures
relaxed over a range of $\kappa$, which approach it from above. The law underestimates $\mathcal D$ by 11 percent for the reference skyrmion and
by 0.2 percent at $R/\Delta\simeq7$, so it is reliable for the large skyrmions where we
use it. The radius alone does not decide the regime of a real film, however, because the
wall width and the damping enter $r$ in the same way.

The pattern must also vary on the length scale that the skyrmion samples. For sinusoidal
coverage with wave number $k=2\pi/\lambda$, period $\lambda$ and first-harmonic
amplitude $f_1$, the odd potential has the
amplitude $U_{\rm amp}=2\pi d\eta|D_0|f_1|\int_0^\infty ds\,\sin^2\theta(s)J_1(ks)|$,
where $J_1$ is a Bessel function. For a narrow wall $\sin^2\theta$ approaches
$\mathrm{sech}^2[(s-R)/\Delta]$, as Fig.~\ref{fig:scaling} shows, and the amplitude
becomes $U_{\rm amp}\simeq4\pi d\eta|D_0|f_1\Delta|J_1(kR)|$. It peaks first at
$kR\simeq1.84$, that is at $\lambda\simeq3.41R$. Much shorter periods are averaged out by
the skyrmion, and much longer ones weaken the coverage gradient. The diffusion response
is therefore nonmonotonic in the period. Figure~\ref{fig:regime}(b) shows
$\Dav/\Dfree$ against the period for a large skyrmion with $R=\SI{322}{nm}$ and
$\Delta=\SI{6}{nm}$, obtained from Eq.~\eqref{eq:tensor} with the finite wall width
retained in the amplitude. For $0.1\leq\alpha\leq0.2$ this skyrmion lies on the
suppressed side, and the suppression is strongest near $\lambda\simeq\SI{1.1}{\micro m}$,
where it reaches 26 percent at $\alpha=0.1$ and 29 percent at $\alpha=0.2$. The strength of the effect is thus tied to a controllable
length scale rather than to an independently chosen barrier height.

The experiments of Ref.~\cite{kapon2025} found that chiral
molecular adsorption can either suppress or enhance
skyrmion diffusion, depending on the magnetic-field
polarity. Our model also permits both regimes, but
through the balance of gyrotropic and dissipative
motion rather than field reversal. Indeed, reversing
the skyrmion charge at fixed potential leaves the
symmetric diffusion tensor unchanged. The observed
field-polarity dependence therefore cannot be
explained by this mechanism alone.

Our results instead identify a distinct contribution
of molecular chirality, a spatially varying odd
interfacial coupling that produces a handed transverse
deflection. We express its strength as the surface DMI coefficient
$D_s=\eta d|D_0|$. Our simulations use $D_s\simeq\SI{1.5}{pJ/m}$, comparable to the
DMI of a heavy-metal interface, which makes the response clear above the thermal noise,
and Fig.~\ref{fig:regime}(b) uses \SI{0.27}{pJ/m}. For weaker molecular couplings the
most sensitive measurements are those that grow linearly with $D_s$. A straight edge of
the molecular coverage running along $\hat{\bm p}$ is the simplest example. The exact
form of Sec.~\ref{sec:symmetry} gives its potential,
$U_{\rm edge}(X)=2D_s\int_{|X|}^\infty ds\,\sin^2\theta\,(1-X^2/s^2)^{1/2}$, a ridge
that extends about $R$ on each side of the edge with a height close to $4D_s\Delta$.
The micromagnetic energy of the reference skyrmion reproduces this ridge within about
one percent at its maximum. In a molecular stripe along $\hat{\bm p}$, one edge
therefore attracts the skyrmion and the other repels it, and the two swap between
enantiomers. The skyrmion spends more time near the attracting edge by the factor
$\exp(2U_{\rm edge}/\kB T)$, whereas changes of anisotropy are the same at both edges
and do not contribute. Edges along $\hat{\bm e}_\parallel$ respond only to the even
in-plane contribution, which the same calculation confirms. A ridge of $0.1\,\kB T$
requires $D_s\simeq\SI{7}{fJ/m}$ for $\Delta=\SI{15}{nm}$, about two hundred times less
than in our simulations. The mean diffusion instead changes quadratically,
$\Dav/\Dfree-1\simeq(r^2-1)(U_{\rm pp}/\kB T)^2/16$ for a weak sinusoidal landscape of
peak-to-trough variation $U_{\rm pp}$, and therefore needs landscapes of order $\kB T$.
The most favorable conditions are thus a film with weak pinning and soft perpendicular
anisotropy, where $\Delta$ is large and micrometer skyrmions can be followed optically,
with molecules adsorbed directly on the magnetic layer in stripes aligned with their
tilt. A periodic molecular pattern of either enantiomer should in addition produce
different diffusion coefficients in the two in-plane directions. Such measurements
could distinguish this mechanism from changes in
anisotropy, pinning, or other adsorption-induced
modifications of the magnetic film.

\section{Conclusions}\label{sec:conclusions}

An aligned chiral adsorbate layer permits an odd interfacial spin coupling, and an
axisymmetric skyrmion responds to its spatial curl. Because the coupling cancels within a
uniform layer, boundaries and coverage variations are the source of translational
forces. Near a localized patch these forces produce a handed transverse deflection, which
we verified with stochastic micromagnetic simulations. In an extended pattern they
suppress diffusion in the direction of the potential variation and enhance it in the
perpendicular direction.

The balance between gyrotropic and dissipative mobilities decides whether this
anisotropy raises or lowers the mean diffusion. The result is exact in one dimension, and
independent calculations on the full two-dimensional landscapes confirm both regimes.
The response persists when the even adsorption anisotropy is varied, and its magnitude
depends on the pattern wavelength through averaging over the skyrmion wall, which offers
a geometric way to strengthen or weaken the coupling. These results establish a
theoretical mechanism for steering and redistributing thermal skyrmion motion. The
handed effects grow linearly with the molecular coupling, whereas the change of diffusion
grows quadratically, so straight edges of the coverage aligned with the molecular tilt
offer the most sensitive test. Transient
deflection and directional diffusion provide complementary observables, both for
controlled patterns and for conditional comparison with molecular experiments.

\begin{acknowledgments}
We thank Artem Volosniev for useful discussions and comments, and Yossi Paltiel for discussions of how the chiral polypeptide
layers are prepared and imaged. The simulations were performed on the scientific
computing clusters of the Institute of Science and Technology Austria and the University of Hamburg. U.P. was funded by the European Union, in the framework of the Marie Sk{\l}odowska-Curie Actions - CISSE Doctoral Network, under Grant Agreement no.\ 101071886. Views and opinions expressed are, however, those of the author(s) only and do not necessarily reflect those of the European Union or the European Research Executive Agency (REA). Neither the European Union nor the granting authority can be held responsible for them.
\end{acknowledgments}

\appendix

\section{Numerical methods and validation}\label{app:numerics}

The reference micromagnetic calculations use a square \SI{1}{nm} mesh with periodic
boundaries. The clean-film and localized-patch trajectories use a \SI{96}{nm} box, and
the small-skyrmion landscape uses a \SI{64}{nm} box containing exactly three coverage
periods. The larger texture is checked on \SI{2}{nm}, \SI{1}{nm} and \SI{0.5}{nm}
meshes, in a \SI{384}{nm} box of eight periods for the two coarser meshes and a
\SI{288}{nm} box of six periods for the finest. Cropping the \SI{1}{nm} texture between
these boxes changes its radius and dissipative integral by less than the quoted
precision. Periodic Gaussian image sums keep the coverage smooth across all boundaries.

At $\alpha=0.4$ the box-size test gives $D/\Dfree=0.75$, $1.04$ and $1.03$ for
\SI{64}{nm}, \SI{96}{nm} and \SI{128}{nm} boxes. The two larger boxes agree with the free
Thiele prediction, and we use \SI{96}{nm} for the baseline. These finite-temperature
diffusivities carry ensemble sampling uncertainty, so we do not use the difference of
two means to infer convergence at the percent level.

Every energy term carries the cell volume $h_xh_yd$, where $h_x$ and $h_y$ are the mesh
spacings. Exchange uses forward differences
and the DM terms use central differences. The effective field is the exact derivative of
this discrete energy, including the gradients of spatially varying coefficients, and
finite-difference checks on random spin configurations agree with it to better than
$2\times10^{-7}$ of the root-mean-square field.

The numerical implementation also contains the even higher-gradient correction
$w_{\rm g}=-(a^2A_{\rm a}/2)[n_x\partial_y^2n_x+n_y\partial_x^2n_y]$, with
the lattice constant $a=\SI{0.3}{nm}$ and $A_{\rm a}=\beta|\Dc|^2/(4A)$, evaluated with discrete periodic
second derivatives in both the energy and the field. We treat it as part of the
numerical model, not as a microscopically required compass interaction. Its contribution
to the potential depth of the larger skyrmion is below $7\times10^{-5}$ on all three
meshes, whereas varying $\beta$ tests the much larger local anisotropy contribution.

We minimize the larger texture in normalized Cartesian spin coordinates using the
tangential projection of the analytic energy gradient, and we record both the optimizer
status and the maximum torque $|\nv\times\Heff|$. The final maximum torques are about
$0.12$, $0.23$ and \SI{1.49}{A/m} on the \SI{2}{nm}, \SI{1}{nm} and \SI{0.5}{nm} meshes.
The corresponding radii are \SI{40.39}{nm}, \SI{43.69}{nm} and \SI{44.75}{nm}, and the
dissipative integrals are $40.46$, $44.55$ and $45.80$. Refining from \SI{1}{nm} to
\SI{0.5}{nm} changes $\Dav/\Dfree$ from $1.257$ to $1.222$ at $\alpha=0.2$ and from
$0.795$ to $0.800$ at $\alpha=0.4$. The crossover is therefore robust, while the numbers
retain a discretization dependence of a few percent.

\begin{figure}[!t]
  \includegraphics[width=\columnwidth]{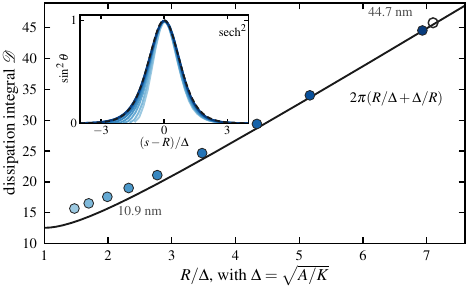}
  \caption{\label{fig:scaling}%
  Test of the circular-wall law used in Sec.~\ref{sec:regime}. Dissipation integral
  $\mathcal D$ against $R/\Delta$ for nine textures relaxed for $0.80\leq\kappa\leq0.99$
  on the \SI{1}{nm} mesh (filled circles) and for $\kappa=0.99$ on the \SI{0.5}{nm} mesh
  (open circle), compared with $2\pi(R/\Delta+\Delta/R)$ (line). Labels give the radii
  of the two transport textures. The inset shows $\sin^2\theta$ against $(s-R)/\Delta$
  for each texture, in the shade of its point, together with
  $\mathrm{sech}^2[(s-R)/\Delta]$ (dashed).}
\end{figure}

Figure~\ref{fig:scaling} tests the circular-wall law of Sec.~\ref{sec:regime}. We
minimize nine textures for $0.80\leq\kappa\leq0.99$ in the same way on the \SI{1}{nm}
mesh, in boxes between 7.9 and 12.8 radii wide, with final maximum torques between
$0.2$ and \SI{1.7}{A/m}. The $\kappa=0.99$ texture reproduces the \SI{1}{nm} radius and
dissipative integral of the larger texture given above. Since $\Delta=\sqrt{A/K}$ is
fixed by the material parameters, the law $2\pi(R/\Delta+\Delta/R)$ contains no free
parameter. It underestimates $\mathcal D$ by 14 percent at $R/\Delta\simeq1.5$ and by
0.2 percent at $R/\Delta\simeq7$, and the wall profiles approach
$\mathrm{sech}^2[(s-R)/\Delta]$ as $R/\Delta$ grows.

We check the thermal-field normalization on independent uniaxial macrospins with
$P(n_z)\propto\exp(KVn_z^2/\kB T)$. The simulated second moments, $0.3879$ and $0.3477$
at \SI{300}{K} and \SI{1000}{K}, differ from the exact values $0.3898$ and $0.3497$ by
less than $0.6$ percent. On the \SI{1}{nm} mesh the stochastic time steps are
\SI{20.75}{fs} for $\alpha\geq0.1$ and \SI{4.15}{fs} at $\alpha=0.02$, where the smaller
step controls the low-damping integration error. At these steps the equilibrium
magnetization of the extended film agrees with the spin-wave estimate within about
$0.5$ percent.

Further checks show that the energy is unchanged, to numerical precision,
under the mirror operation combined with the axial spin transformation and enantiomer
exchange. A nonuniform curl-free test field $\Dc=(D_*\cos kx,0)$, with amplitude $D_*$ and
wave number $k$, produces a potential
variation below $10^{-9}\kB T$ for an axisymmetric texture in a sufficiently large box.
FFT correlation of the discrete coupling fields with the texture densities reproduces
explicit rigid translations to the same precision.

\section{Position and displacement statistics}\label{app:centre}

The position estimator weights the coordinate by the signed topological density $q$
through a circular mean that respects the periodic boundaries,
\begin{equation}
 X_{\rm c}=\frac{L_x}{2\pi}\arg
 \sum_{\bm r}\frac{q(\bm r)}{\sum_{\bm r'}q(\bm r')}
 e^{2\pi i x/L_x},
 \label{eq:circmean}
\end{equation}
and similarly for $Y_{\rm c}$, where $x$ is the coordinate of cell $\bm r$ and $L_x$ is
the box length along $x$. Successive
increments are unwrapped before displacements are accumulated. Normalizing by the signed
charge removes the extensive positive thermal background that biases positive-weight
estimators. In tests with component noise of amplitude $0.1$, the position bias is about
\SI{0.024}{nm}, compared with \SI{0.208}{nm} for a positive-weight estimator. The
circular estimator also passes tests with the skyrmion placed on the periodic boundary.
These tests characterize its numerical accuracy and do not prove exact unbiasedness at
arbitrary temperature.

\label{app:msd}
For the free-film trajectories we average displacement variances over time origins and
independent trajectories, using the connected second moment
$C(\tau)=\langle|\Delta\bm R(\tau)-\langle\Delta\bm R(\tau)\rangle|^2\rangle$ of the
displacement $\Delta\bm R(\tau)$ over a lag $\tau$, because a
mean drift would otherwise add a contribution quadratic in lag. Single-origin ensemble
averages have the same expectation for stationary Brownian increments and are not biased
downward, but they have a larger sampling uncertainty.

The stochastic collective-coordinate estimates use uncentered time-averaged second
moments, with equilibrium initial conditions and zero mean drift expected by symmetry.
We fit them through the origin over lags up to one fifth of the trajectory duration and
use running diffusivities as a finite-lag diagnostic. Bootstrap resampling over whole
trajectories retains the correlations between time origins. These estimates serve as
independent checks of the asymptotic cell calculation of Appendix~\ref{app:cell} rather
than defining its long-time limit.

Figures~\ref{fig:drift}(a)--\ref{fig:drift}(c) instead report mean displacements with
standard errors computed separately for each sign. The two signs share thermal-noise
seeds, which reduces the variance of their difference but does not cancel fluctuations
exactly in nonlinear dynamics. The separate errors are therefore conservative for the
difference.

\section{Periodic transport calculation}\label{app:cell}

An independent homogenization calculation determines the long-time diffusion from the
periodic landscape without fitting trajectories. We write $u=U/\kB T$ and
$g=G/(\alpha\mathcal D)$ and choose the unit of time so that $\Dfree=1$. The drift is
then $\bm b=-\nabla u+g\,\zh\times\nabla u$, and the stationary density is
$\rho=e^{-u}/\int e^{-u}$. The periodic correctors $\Psi_i$, one for each
direction $i=x,y$, obey
$(\nabla^2+\bm b\cdot\nabla)\Psi_i=-b_i$ with zero spatial mean and give the effective
symmetric tensor
\begin{equation}
 \frac{D_{ij}}{\Dfree}
 =\left\langle
 (\hat{\bm e}_i+\nabla\Psi_i)\cdot
 (\hat{\bm e}_j+\nabla\Psi_j)
 \right\rangle_\rho ,
 \label{eq:cell}
\end{equation}
where $\hat{\bm e}_i$ are the Cartesian unit vectors and $\langle\cdot\rangle_\rho$ is
the average over one cell weighted by $\rho$. We solve for the correctors with Fourier
derivatives and a preconditioned iterative
linear solver, check the residual, and refine the periodic grid independently of the
micromagnetic mesh.

For a one-dimensional potential we define the normalized weight
$h=e^u/\langle e^u\rangle$, with a uniform average over one period. The corrector
derivatives are then $\Psi_x'=h-1$ and $\Psi_y'=-g(h-1)$, and the identities
$\langle h\rangle_\rho=\langle h^2\rangle_\rho=L$ yield Eq.~\eqref{eq:tensor}, including
its vanishing off-diagonal component. The normalization of $h$ is essential for this
result.

The solver reproduces the exact tensor for a cosine potential to better than $10^{-8}$
in both dissipative and gyrotropic test cases. The coefficients of the small-skyrmion
landscape are unchanged on $48^2$, $72^2$ and $96^2$ grids, those of the larger
landscape change by less than $10^{-7}$ on $64^2$, $96^2$ and $128^2$ grids, and typical
relative residuals of the linear system are below $2\times10^{-11}$. These tolerances
concern only the reduced transport calculation. They are independent of the
micromagnetic mesh error and of the rigid-texture approximation.

\bibliography{refs}

@article{rieger2023,
  author = {Rieger, Philipp and Wei{\ss}enhofer, Markus and Nowak, Ulrich},
  title = {Defect-enhanced diffusion of magnetic skyrmions},
  journal = {Phys. Rev. B}, volume = {108}, pages = {144417}, year = {2023},
  doi = {10.1103/PhysRevB.108.144417}
}

@article{schick2024,
  author = {Schick, Daniel and Wei{\ss}enhofer, Markus and R{\'o}zsa, Levente and
            Roth{\"o}rl, Jan and Virnau, Peter and Nowak, Ulrich},
  title = {Two levels of topology in skyrmion lattice dynamics},
  journal = {Phys. Rev. Research}, volume = {6}, pages = {013097}, year = {2024},
  doi = {10.1103/PhysRevResearch.6.013097}
}

@article{kuliashov2026,
  title   = {Supercurrents in {Josephson} junctions with chiral molecular potentials},
  author  = {Kuliashov, Oleg and Cappellaro, Alberto and Millo, Oded and Paltiel, Yossi and Lemeshko, Mikhail and Alhyder, Ragheed},
  journal = {Physical Review B},
  volume  = {114},
  number  = {15},
  pages   = {154510},
  year    = {2026},
  doi     = {10.1103/S1JN-FQ4D}
}

@article{alhyder2025,
  title   = {Quantum transport in the presence of a chiral molecular potential},
  author  = {Alhyder, Ragheed and Lemeshko, Mikhail and Cappellaro, Alberto},
  journal = {The Journal of Chemical Physics},
  volume  = {162},
  number  = {23},
  pages   = {234106},
  year    = {2025},
  doi     = {10.1063/5.0271155}
}

@article{alhyder2023,
  title   = {Achiral dipoles on a ferromagnet can affect its magnetization direction},
  author  = {Alhyder, Ragheed and Cappellaro, Alberto and Lemeshko, Mikhail and Volosniev, Artem},
  journal = {The Journal of Chemical Physics},
  volume  = {159},
  number  = {10},
  pages   = {104103},
  year    = {2023},
  doi     = {10.1063/5.0165806}
}

@article{skyrme1961,
  author = {Skyrme, T. H. R.},
  title = {A non-linear field theory},
  journal = {Proc. R. Soc. Lond. A}, volume = {260}, pages = {127}, year = {1961},
  doi = {10.1098/rspa.1961.0018}
}

@article{bogdanov1989,
  author = {Bogdanov, A. N. and Yablonskii, D. A.},
  title = {Thermodynamically stable ``vortices'' in magnetically ordered crystals},
  journal = {Sov. Phys. JETP}, volume = {68}, pages = {101}, year = {1989}
}

@article{Dohi2022,
  title = {Thin Film Skyrmionics},
  volume = {13},
  ISSN = {1947-5462},
  url = {http://dx.doi.org/10.1146/annurev-conmatphys-031620-110344},
  DOI = {10.1146/annurev-conmatphys-031620-110344},
  number = {1},
  journal = {Annual Review of Condensed Matter Physics},
  publisher = {Annual Reviews},
  author = {Dohi,  Takaaki and Reeve,  Robert M. and Kl\"{a}ui,  Mathias},
  year = {2022},
  month = Mar,
  pages = {73–95}
}

@article{Finocchio2016,
  title = {Magnetic skyrmions: from fundamental to applications},
  volume = {49},
  ISSN = {1361-6463},
  url = {http://dx.doi.org/10.1088/0022-3727/49/42/423001},
  DOI = {10.1088/0022-3727/49/42/423001},
  number = {42},
  journal = {Journal of Physics D: Applied Physics},
  publisher = {IOP Publishing},
  author = {Finocchio,  Giovanni and B\"{u}ttner,  Felix and Tomasello,  Riccardo and Carpentieri,  Mario and Kl\"{a}ui,  Mathias},
  year = {2016},
  month = Sep,
  pages = {423001}
}

@article{Mishra2025,
  title = {Magnetic skyrmion: from fundamental physics to pioneering applications},
  volume = {12},
  ISSN = {1931-9401},
  url = {http://dx.doi.org/10.1063/5.0223004},
  DOI = {10.1063/5.0223004},
  number = {1},
  journal = {Applied Physics Reviews},
  publisher = {AIP Publishing},
  author = {Mishra,  Kishan K. and Lone,  Aijaz H. and Srinivasan,  Srikant and Fariborzi,  Hossein and Setti,  Gianluca},
  year = {2025},
  month = Feb, 
    pages={011315}
}

@article{muhlbauer2009,
  author = {M\"uhlbauer, S. and Binz, B. and Jonietz, F. and Pfleiderer, C. and
            Rosch, A. and Neubauer, A. and Georgii, R. and B\"oni, P.},
  title = {Skyrmion lattice in a chiral magnet},
  journal = {Science}, volume = {323}, pages = {915}, year = {2009},
  doi = {10.1126/science.1166767}
}

@article{yu2010,
  author = {Yu, X. Z. and Onose, Y. and Kanazawa, N. and Park, J. H. and Han, J. H.
            and Matsui, Y. and Nagaosa, N. and Tokura, Y.},
  title = {Real-space observation of a two-dimensional skyrmion crystal},
  journal = {Nature (London)}, volume = {465}, pages = {901}, year = {2010},
  doi = {10.1038/nature09124}
}

@article{heinze2011,
  author = {Heinze, Stefan and von Bergmann, Kirsten and Menzel, Matthias and
            Brede, Jens and Kubetzka, Andr\'e and Wiesendanger, Roland and
            Bihlmayer, Gustav and Bl\"ugel, Stefan},
  title = {Spontaneous atomic-scale magnetic skyrmion lattice in two dimensions},
  journal = {Nat. Phys.}, volume = {7}, pages = {713}, year = {2011},
  doi = {10.1038/nphys2045}
}

@article{fert2013,
  author = {Fert, Albert and Cros, Vincent and Sampaio, Jo{\~a}o},
  title = {Skyrmions on the track},
  journal = {Nat. Nanotechnol.}, volume = {8}, pages = {152}, year = {2013},
  doi = {10.1038/nnano.2013.29}
}

@article{nagaosa2013,
  author = {Nagaosa, Naoto and Tokura, Yoshinori},
  title = {Topological properties and dynamics of magnetic skyrmions},
  journal = {Nat. Nanotechnol.}, volume = {8}, pages = {899}, year = {2013},
  doi = {10.1038/nnano.2013.243}
}

@article{fert2017,
  author = {Fert, Albert and Reyren, Nicolas and Cros, Vincent},
  title = {Magnetic skyrmions: advances in physics and potential applications},
  journal = {Nat. Rev. Mater.}, volume = {2}, pages = {17031}, year = {2017},
  doi = {10.1038/natrevmats.2017.31}
}

@article{back2020,
  author = {Back, C. and Cros, V. and Ebert, H. and Everschor-Sitte, K. and Fert, A.
            and Garst, M. and Ma, Tianping and Mankovsky, S. and Monchesky, T. L.
            and Mostovoy, M. and Nagaosa, N. and Parkin, S. S. P. and Pfleiderer, C.
            and Reyren, N. and Rosch, A. and Taguchi, Y. and Tokura, Y. and
            von Bergmann, K. and Zang, Jiadong},
  title = {The 2020 skyrmionics roadmap},
  journal = {J. Phys. D: Appl. Phys.}, volume = {53}, pages = {363001}, year = {2020},
  doi = {10.1088/1361-6463/ab8418}
}

@article{sampaio2013,
  author = {Sampaio, Jo{\~a}o and Cros, Vincent and Rohart, Stanislas and
            Thiaville, Andr\'e and Fert, Albert},
  title = {Nucleation, stability and current-induced motion of isolated magnetic
           skyrmions in nanostructures},
  journal = {Nat. Nanotechnol.}, volume = {8}, pages = {839--844}, year = {2013},
  doi = {10.1038/nnano.2013.210}
}

@article{woo2016,
  author = {Woo, Seonghoon and Litzius, Kai and Kr\"uger, Benjamin and Im, Mi-Young
            and Caretta, Lucas and Richter, Kornel and Mann, Maxwell and Krone, Axel
            and Reeve, Robert M. and Weigand, Markus and Agrawal, Parnika and
            Lemesh, Ivan and Mawass, Mohamed-Assaad and Fischer, Peter and
            Kl\"aui, Mathias and Beach, Geoffrey S. D.},
  title = {Observation of room-temperature magnetic skyrmions and their
           current-driven dynamics in ultrathin metallic ferromagnets},
  journal = {Nat. Mater.}, volume = {15}, pages = {501--506}, year = {2016},
  doi = {10.1038/nmat4593}
}

@article{moreauluchaire2016,
  author = {Moreau-Luchaire, C. and Moutafis, C. and Reyren, N. and Sampaio, J.
            and Vaz, C. A. F. and Van Horne, N. and Bouzehouane, K. and Garcia, K.
            and Deranlot, C. and Warnicke, P. and Wohlh\"uter, P. and George, J.-M.
            and Weigand, M. and Raabe, J. and Cros, V. and Fert, A.},
  title = {Additive interfacial chiral interaction in multilayers for stabilization
           of small individual skyrmions at room temperature},
  journal = {Nat. Nanotechnol.}, volume = {11}, pages = {444--448}, year = {2016},
  doi = {10.1038/nnano.2015.313}
}

@article{boulle2016,
  author = {Boulle, O. and Vogel, J. and Yang, H. and Pizzini, S. and
            de Souza Chaves, D. and Locatelli, A. and Mente\c{s}, T. O. and Sala, A.
            and Aballe, L. and Foerster, M. and Belmeguenai, M. and Roussign\'e, Y.
            and Stashkevich, A. and Ch\'erif, S. M. and Buda-Prejbeanu, L. D. and
            Klein, O. and Chshiev, M. and Auffret, S. and Miron, I. M. and Gaudin, G.},
  title = {Room-temperature chiral magnetic skyrmions in ultrathin magnetic
           nanostructures},
  journal = {Nat. Nanotechnol.}, volume = {11}, pages = {449--454}, year = {2016},
  doi = {10.1038/nnano.2015.315}
}

@article{thiele1973,
  author = {Thiele, A. A.},
  title = {Steady-state motion of magnetic domains},
  journal = {Phys. Rev. Lett.}, volume = {30}, pages = {230}, year = {1973},
  doi = {10.1103/PhysRevLett.30.230}
}

@article{everschor2012,
  author = {Everschor, Karin and Garst, Markus and Binz, Benedikt and Jonietz, Florian
            and M\"uhlbauer, Sebastian and Pfleiderer, Christian and Rosch, Achim},
  title = {Rotating skyrmion lattices by spin torques and field or temperature
           gradients},
  journal = {Phys. Rev. B}, volume = {86}, pages = {054432}, year = {2012},
  doi = {10.1103/PhysRevB.86.054432}
}

@article{everschor2018,
  author = {Everschor-Sitte, Karin and Masell, Jan and Reeve, Rebecca M. and
            Kl\"aui, Mathias},
  title = {Perspective: Magnetic skyrmions---Overview of recent progress in an
           active research field},
  journal = {J. Appl. Phys.}, volume = {124}, pages = {240901}, year = {2018},
  doi = {10.1063/1.5048972}
}

@article{iwasaki2013,
  author = {Iwasaki, Junichi and Mochizuki, Masahito and Nagaosa, Naoto},
  title = {Universal current-velocity relation of skyrmion motion in chiral magnets},
  journal = {Nat. Commun.}, volume = {4}, pages = {1463}, year = {2013},
  doi = {10.1038/ncomms2442}
}

@article{makhfudz2012,
  author = {Makhfudz, Imam and Kr\"uger, Benjamin and Tchernyshyov, Oleg},
  title = {Inertia and chiral edge modes of a skyrmion magnetic bubble},
  journal = {Phys. Rev. Lett.}, volume = {109}, pages = {217201}, year = {2012},
  doi = {10.1103/PhysRevLett.109.217201}
}

@article{schutte2014,
  author = {Sch\"utte, Christoph and Iwasaki, Junichi and Rosch, Achim and
            Nagaosa, Naoto},
  title = {Inertia, diffusion, and dynamics of a driven skyrmion},
  journal = {Phys. Rev. B}, volume = {90}, pages = {174434}, year = {2014},
  doi = {10.1103/PhysRevB.90.174434}
}

@article{jiang2017,
  author = {Jiang, Wanjun and Zhang, Xichao and Yu, Guoqiang and Zhang, Wei and
            Wang, Xiao and Benjamin Jungfleisch, M. and Pearson, John E. and
            Cheng, Xuemei and Heinonen, Olle and Wang, Kang L. and Zhou, Yan and
            Hoffmann, Axel and te Velthuis, Suzanne G. E.},
  title = {Direct observation of the skyrmion {H}all effect},
  journal = {Nat. Phys.}, volume = {13}, pages = {162}, year = {2017},
  doi = {10.1038/nphys3883}
}

@article{litzius2017,
  author = {Litzius, Kai and Lemesh, Ivan and Kr\"uger, Benjamin and Bassirian, Pedram
            and Caretta, Lucas and Richter, Kornel and B\"uttner, Felix and Sato, Koji
            and Tretiakov, Oleg A. and F\"orster, Johannes and Reeve, Robert M. and
            Weigand, Markus and Bykova, Iuliia and Stoll, Hermann and Sch\"utz, Gisela
            and Beach, Geoffrey S. D. and Kl\"aui, Mathias},
  title = {Skyrmion {H}all effect revealed by direct time-resolved {X}-ray microscopy},
  journal = {Nat. Phys.}, volume = {13}, pages = {170--175}, year = {2017},
  doi = {10.1038/nphys4000}
}

@article{brown1963,
  author = {Brown, William Fuller},
  title = {Thermal fluctuations of a single-domain particle},
  journal = {Phys. Rev.}, volume = {130}, pages = {1677}, year = {1963},
  doi = {10.1103/PhysRev.130.1677}
}

@article{kubo1966,
  author = {Kubo, R.},
  title = {The fluctuation-dissipation theorem},
  journal = {Rep. Prog. Phys.}, volume = {29}, pages = {255}, year = {1966},
  doi = {10.1088/0034-4885/29/1/306}
}

@article{garciapalacios1998,
  author = {Garc\'ia-Palacios, Jos\'e Luis and L\'azaro, Francisco J.},
  title = {Langevin-dynamics study of the dynamical properties of small magnetic
           particles},
  journal = {Phys. Rev. B}, volume = {58}, pages = {14937}, year = {1998},
  doi = {10.1103/PhysRevB.58.14937}
}

@article{zazvorka2019,
  author = {Z\'azvorka, Jakub and Jakobs, Florian and Heinze, Daniel and Keil, Niklas
            and Kromin, Sascha and Jaiswal, Samridh and Litzius, Kai and
            Jakob, Gerhard and Virnau, Peter and Pinna, Daniele and
            Everschor-Sitte, Karin and R\'ozsa, Levente and Donges, Andreas and
            Nowak, Ulrich and Kl\"aui, Mathias},
  title = {Thermal skyrmion diffusion used in a reshuffler device},
  journal = {Nat. Nanotechnol.}, volume = {14}, pages = {658}, year = {2019},
  doi = {10.1038/s41565-019-0436-8}
}

@article{miltat2018,
  author = {Miltat, Jacques and Rohart, Stanislas and Thiaville, Andr\'e},
  title = {Brownian motion of magnetic domain walls and skyrmions, and their
           diffusion constants},
  journal = {Phys. Rev. B}, volume = {97}, pages = {214426}, year = {2018},
  doi = {10.1103/PhysRevB.97.214426}
}

@article{nozaki2019,
  author = {Nozaki, Takayuki and Jibiki, Yuma and Goto, Minori and Tamura, Eiiti and
            Nozaki, Tatsuya and Kubota, Hitoshi and Fukushima, Akio and
            Yuasa, Shinji and Suzuki, Yoshishige},
  title = {Brownian motion of skyrmion bubbles and its control by voltage
           applications},
  journal = {Appl. Phys. Lett.}, volume = {114}, pages = {012402}, year = {2019},
  doi = {10.1063/1.5070101}
}

@article{dohi2023,
  author = {Dohi, Takaaki and Wei{\ss}enhofer, Markus and Kerber, Nico and
            Kammerbauer, Fabian and Ge, Yuqing and Raab, Klaus and Z\'azvorka, Jakub
            and Syskaki, Maria-Andromachi and Shahee, Aga and Ruhwedel, Moritz and
            B\"ottcher, Tobias and Pirro, Philipp and Jakob, Gerhard and Nowak, Ulrich
            and Kl\"aui, Mathias},
  title = {Enhanced thermally-activated skyrmion diffusion with tunable effective
           gyrotropic force},
  journal = {Nat. Commun.}, volume = {14}, pages = {5424}, year = {2023},
  doi = {10.1038/s41467-023-40720-0}
}

@article{weissenhofer2020,
  author = {Wei{\ss}enhofer, Markus and Nowak, Ulrich},
  title = {Diffusion of skyrmions: the role of topology and anisotropy},
  journal = {New J. Phys.}, volume = {22}, pages = {103059}, year = {2020},
  doi = {10.1088/1367-2630/abc1c9}
}

@article{zhao2020,
  author = {Zhao, Le and Wang, Zidong and Zhang, Xichao and Liang, Xue and Xia, Jing
            and Wu, Keyu and Zhou, Heng-An and Dong, Yiqing and Yu, Guoqiang and
            Wang, Kang L. and Liu, Xiaoxi and Zhou, Yan and Jiang, Wanjun},
  title = {Topology-dependent Brownian gyromotion of a single skyrmion},
  journal = {Phys. Rev. Lett.}, volume = {125}, pages = {027206}, year = {2020},
  doi = {10.1103/PhysRevLett.125.027206}
}

@article{prychynenko2018,
  author = {Prychynenko, Diana and Sitte, Matthias and Litzius, Kai and
            Kr\"uger, Benjamin and Bourianoff, George and Kl\"aui, Mathias and
            Sinova, Jairo and Everschor-Sitte, Karin},
  title = {Magnetic skyrmion as a nonlinear resistive element: A potential building
           block for reservoir computing},
  journal = {Phys. Rev. Applied}, volume = {9}, pages = {014034}, year = {2018},
  doi = {10.1103/PhysRevApplied.9.014034}
}

@article{pinna2018,
  author = {Pinna, Daniele and Abreu Araujo, Flavio and Kim, Joo-Von and
            Cros, Vincent and Querlioz, Damien and Bessiere, Pierre and
            Droulez, Jacques and Grollier, Julie},
  title = {Skyrmion gas manipulation for probabilistic computing},
  journal = {Phys. Rev. Applied}, volume = {9}, pages = {064018}, year = {2018},
  doi = {10.1103/PhysRevApplied.9.064018}
}

@article{raab2022,
  author = {Raab, Klaus and Brems, Maarten A. and Beneke, Grischa and Dohi, Takaaki
            and Roth\"orl, Jan and Kammerbauer, Fabian and Mentink, Johan H. and
            Kl\"aui, Mathias},
  title = {Brownian reservoir computing realized using geometrically confined
           skyrmion dynamics},
  journal = {Nat. Commun.}, volume = {13}, pages = {6982}, year = {2022},
  doi = {10.1038/s41467-022-34309-2}
}

@article{crepieux1998,
  author = {Cr\'epieux, A. and Lacroix, C.},
  title = {Dzyaloshinsky-{M}oriya interactions induced by symmetry breaking at a
           surface},
  journal = {J. Magn. Magn. Mater.}, volume = {182}, pages = {341}, year = {1998},
  doi = {10.1016/S0304-8853(97)01044-5}
}

@article{kim2013,
  author = {Kim, Kyoung-Whan and Lee, Hyun-Woo and Lee, Kyung-Jin and
            Stiles, M. D.},
  title = {Chirality from interfacial spin-orbit coupling effects in magnetic
           bilayers},
  journal = {Phys. Rev. Lett.}, volume = {111}, pages = {216601}, year = {2013},
  doi = {10.1103/PhysRevLett.111.216601}
}

@article{kundu2015,
  author = {Kundu, Anirban and Zhang, Shufeng},
  title = {Dzyaloshinskii-{M}oriya interaction mediated by spin-polarized band with
           {R}ashba spin-orbit coupling},
  journal = {Phys. Rev. B}, volume = {92}, pages = {094434}, year = {2015},
  doi = {10.1103/PhysRevB.92.094434}
}

@article{nembach2015,
  author = {Nembach, Hans T. and Shaw, Justin M. and Weiler, Mathias and
            Ju\'e, Emilie and Silva, Thomas J.},
  title = {Linear relation between {H}eisenberg exchange and interfacial
           {D}zyaloshinskii-{M}oriya interaction in metal films},
  journal = {Nat. Phys.}, volume = {11}, pages = {825}, year = {2015},
  doi = {10.1038/nphys3418}
}

@article{srivastava2018,
  author = {Srivastava, Titiksha and Schott, Marine and Juge, Roméo and
            K\v{r}i\v{z}\'akov\'a, Viola and Belmeguenai, Mohamed and Roussign\'e, Yves
            and Bernand-Mantel, Anne and Ranno, Laurent and Pizzini, Stefania and
            Ch\'erif, Salim-Mourad and Stashkevich, Andrey and Auffret, Stéphane and
            Boulle, Olivier and Gaudin, Gilles and Chshiev, Mairbek and
            Baraduc, Claire and B\'ea, H\'el\`ene},
  title = {Large-voltage tuning of {D}zyaloshinskii-{M}oriya interactions: A route
           toward dynamic control of skyrmion chirality},
  journal = {Nano Lett.}, volume = {18}, pages = {4871}, year = {2018},
  doi = {10.1021/acs.nanolett.8b01502}
}

@article{rohart2013,
  author = {Rohart, Stanislas and Thiaville, Andr\'e},
  title = {Skyrmion confinement in ultrathin film nanostructures in the presence of
           {D}zyaloshinskii--{M}oriya interaction},
  journal = {Phys. Rev. B}, volume = {88}, pages = {184422}, year = {2013},
  doi = {10.1103/PhysRevB.88.184422}
}

@article{gohler2011,
  author = {G\"ohler, B. and Hamelbeck, V. and Markus, T. Z. and Kettner, M. and
            Hanne, G. F. and Vager, Z. and Naaman, R. and Zacharias, H.},
  title = {Spin selectivity in electron transmission through self-assembled
           monolayers of double-stranded {DNA}},
  journal = {Science}, volume = {331}, pages = {894}, year = {2011},
  doi = {10.1126/science.1199339}
}

@article{naaman2012,
  author = {Naaman, Ron and Waldeck, David H.},
  title = {Chiral-induced spin selectivity effect},
  journal = {J. Phys. Chem. Lett.}, volume = {3}, pages = {2178}, year = {2012},
  doi = {10.1021/jz300793y}
}

@article{naaman2019,
  author = {Naaman, Ron and Paltiel, Yossi and Waldeck, David H.},
  title = {Chiral molecules and the electron spin},
  journal = {Nat. Rev. Chem.}, volume = {3}, pages = {250}, year = {2019},
  doi = {10.1038/s41570-019-0087-1}
}

@article{bendor2017,
  author = {Ben Dor, Oren and Yochelis, Shira and Radko, Anna and Vankayala, Kiran
            and Capua, Eyal and Capua, Amir and Yang, See-Hun and Baczewski,
            Lech Tomasz and Parkin, Stuart Stephen Papworth and Naaman, Ron and
            Paltiel, Yossi},
  title = {Magnetization switching in ferromagnets by adsorbed chiral molecules
           without current or external magnetic field},
  journal = {Nat. Commun.}, volume = {8}, pages = {14567}, year = {2017},
  doi = {10.1038/ncomms14567}
}

@article{kapon2025,
  author = {Kapon, Yael and Kammerbauer, Fabian and Balland, Theo and
            Yochelis, Shira and Kl\"aui, Mathias and Paltiel, Yossi},
  title = {Effects of chiral polypeptides on skyrmion stability and dynamics},
  journal = {Nano Lett.}, volume = {25}, pages = {306}, year = {2025},
  doi = {10.1021/acs.nanolett.4c05035}
}

@article{bloom2024,
  author = {Bloom, Brian P. and Paltiel, Yossi and Naaman, Ron and
            Waldeck, David H.},
  title = {Chiral induced spin selectivity},
  journal = {Chem. Rev.}, volume = {124}, pages = {1950--1991}, year = {2024},
  doi = {10.1021/acs.chemrev.3c00661}
}

@article{mentink2010,
  author = {Mentink, J. H. and Tretyakov, M. V. and Fasolino, A. and
            Katsnelson, M. I. and Rasing, Th.},
  title = {Stable and fast semi-implicit integration of the stochastic
           {L}andau-{L}ifshitz equation},
  journal = {J. Phys.: Condens. Matter}, volume = {22}, pages = {176001}, year = {2010},
  doi = {10.1088/0953-8984/22/17/176001}
}

@article{lifson1962,
  title = {Self-Diffusion of Ions in Polyelectrolyte Solutions},
  author = {Lifson, S. and Jackson, J. L.},
  journal = {J. Chem. Phys.},
  volume = {36},
  pages = {2410},
  year = {1962},
  doi = {10.1063/1.1732899}
}

@article{hondamuni2026,
  author = {Hondamuni, Buddhika and Balland, Th{\'e}o and Kammerbauer, Fabian and
            Moharana, Ashish and Bindu and Singh, Amandeep and Ozeri, Meital and
            Yochelis, Shira and Paltiel, Yossi and Dhungel, Omkar and Kazi, Zeeshawn and
            Fu, Kai-Mei C. and Watanabe, Hideyuki and Kl{\"a}ui, Mathias and
            Wickenbrock, Arne and Bar-Gill, Nir and Wittmann, Angela and Budker, Dmitry},
  title = {Magnetic microscopy of skyrmions in magnetic thin films with chiral overlayers},
  journal = {Phys. Rev. Appl.}, volume = {26}, pages = {044075}, year = {2026},
  doi = {10.1103/thmz-53f2}
}

\end{document}